\documentclass[twocolumn,aps,showpacs]{revtex4}
\usepackage{amsmath,amssymb}
\usepackage{epsfig}

\begin{document}

\title{Long-range correlated stationary Markovian processes}

\author{Fabrizio Lillo$^{*}$,
        Salvatore Miccich\`e$^{*,\dag}$ 
        and Rosario N. Mantegna $^{*,\dag}$}

\affiliation{$^{*}$ Istituto Nazionale per la Fisica della Materia, Unit\`a di Palermo
                    Viale delle Scienze, I-90128 Palermo, Italy \\ 
             $^{\dag}$ Dipartimento di Fisica e Tecnologie Relative, Universit\`a di Palermo, 
                       Viale delle Scienze, I-90128 Palermo, Italy}

\begin{abstract}
We introduce a new class of stochastic processes which are stationary, Markovian and characterized by an infinite range 
of time-scales. By transforming the Fokker-Planck equation of the process into a Schr\"odinger equation with an appropriate quantum potential we determine the asymptotic behavior of the autocorrelation function of the process in an analytical way. We find the conditions needed to observe a stationary long-range correlated Markovian process.
In the presence of long-range correlation, for selected values of the control parameters, the process has a $1/f$-like spectral density for low frequency values.
\end{abstract}

\pacs{02.50.Ey, 05.10.Gg, 05.40.-a, 02.50.Ga}

\maketitle

Since the classical works on Brownian motion \cite{BM,ornstein} 
a great variety of physical systems has been modeled in terms of stochastic
processes \cite{VanKampen81}. Stochastic processes can be differently classified
depending on the properties of their
conditional probability densities. Among random processes, 
Markov processes play a central role in the modeling of 
natural phenomena. A process $x(t)$ is said to be a Markov process
if the conditional probability density $P(x_{n+1},t_{n+1}|x_{n},t_{n}; \dots ;x_{1},t_{1})$ depends only on the
last value $x_n$ at $t_n$ and not on the previous values $x_{n-1}$ at $t_{n-1}$,
$x_{n-2}$ at $t_{n-2}$, etc. The transition probability of any Markov
process fulfills the Chapman-Kolmogorov equation \cite{VanKampen81}.
It is worth noting that  a Markov process is fully determined 
by the knowledge of the probability density function (pdf)
$W(x,t)$ of the process and the transition probability
$P(x_{n+1},t_{n+1}|x_{n},t_{n})$. Such level of simplicity
is rather unique among stochastic processes. In fact, a 
non-Markovian process is characterized by an infinite hierarchy of 
transition probabilities. When the Markovian process is continuous
both in space and time, the time evolution of the pdf is described by a Fokker-Planck (FP)
equation. 

Another classification of stochastic processes 
considers the nature of correlation
of the random variable. Under this classification,
random variables are divided in short-range and 
long-range correlated variables. Short-range 
correlated variables are characterized by a finite mean of time-scales 
of the process whereas a similar mean time-scale 
does not exist for long-range correlated variables. 
An equivalent definition can be 
given by considering the finiteness or infiniteness of the 
integral of the autocorrelation function of the random process
\cite{Cassandro78,Bouchaud90,Samorodnitsky94}.
Long-range correlated processes $x(t)$ are usually characterized 
by a $1/f$-like spectral density and $1/f$ noise
has been observed in several physical systems \cite{Press78,Weissman88}.
In the presence of long-range correlation, the time integral $s(t)$ of 
the process $x(t)$ is a superdiffusive 
stochastic process showing 
$\langle \Delta s^2(t)\rangle \sim D_\gamma\,t^{\gamma}$ with $\gamma>1$ and $D_\gamma$ is a constant.
Superdiffusive stochastic processes have been observed in
several physical systems. A classical example 
is Richardson's observation that two particles moving
in a turbulent fluid which at time $t=0$ are originally placed 
very close the one with the other have a relative separation 
$\ell$ at time $t$ that follows the relation $<\ell^2(t)> \propto t^3$ \cite{Richardson26}. 
More recent examples include anomalous kinetic in chaotic dynamics 
due to flights and trapping \cite{Geisel85}, dynamics
of aggregate of amphiphilic molecules \cite{Ott90} and 
dynamics of a tracer in a two-dimensional rotating flow 
\cite{Solomon93}. Several non-Markovian \cite{Kubo,Mandelbrot,Klafter} or non-stationary \cite{Procaccia,drift} 
models of long-range correlated and anomalous 
diffusing processes have been developed.
  
Stationary Markovian processes are usually short-range correlated. However, 
the formal definition of a Markovian process does not imply that
all stationary Markovian processes must be short-range correlated.
In this Letter, we analytically show that there exists a class of 
stationary Markovian processes with an asymptotically power-law decaying autocorrelation function. 
For a finite range of the control parameters the process is long-range correlated. 
To this end, let us consider
a continuous Markovian stochastic process $x(t)$
whose pdf $W(x,t)$ is described by
the FP equation with constant diffusion coefficient
$\partial_t W=-\partial_x(D^{(1)}(x)W)+D\,\partial^2_x W$. For the sake of simplicity, in this study we set $D=1$.
The eigenvalue spectrum of the FP equation describing a stationary process
consists of a discrete part 
${\lambda_0=0,\lambda_1,...,\lambda_p}$ and a continuous part 
$]\lambda_c,+\infty[$  ($\lambda_c \geqslant \lambda_p$) 
associated with eigenfunctions $\varphi_{\lambda}$. The stationary
pdf is $W(x)=\varphi_{0}$. The FP equation 
with constant diffusion coefficient can be transformed into a Schr\"odinger 
equation \cite{risken} with a quantum potential 
$V_S(x)=(D^{(1)}(x))^2/4+\partial_x D^{(1)}(x)/2$.
The eigenvalue spectrum of the Schr\"odinger 
equation is equal to the eigenvalue spectrum of the
FP equation. The relation between the 
eigenfunctions of the FP equation and the 
eigenfunctions $\psi_{\lambda}$ of the Schr\"odinger equation
is $\varphi_{\lambda}=\psi_{\lambda}\psi_0$. For a stationary process the 2-point probability density function 
$W_2(x,t;x',t+\tau)$ can be expressed in terms of 
the eigenfunctions of the Schr\"odinger equation. Specifically, one can write
\begin{eqnarray}
           &&   \hspace{-0.15 in}
                W_2(x,t;x',t+\tau)~=~\psi_0(x)~\psi_0(x') \times \label{W2} \\
           &&   \hspace{-0.1 in}\left( 
                                    \sum_{\lambda=\lambda_1}^{\lambda_p}\,\psi_\lambda(x)\,\psi_\lambda(x')\,e^{- \lambda \tau}\! + \! 
                                    \int_{\lambda_c}^{+\infty} \! d \lambda\,\psi_\lambda(x)\,\psi_\lambda(x')\,e^{- \lambda \tau}
                                \right) .      \nonumber
\end{eqnarray}
Eq. $(\ref{W2})$ extends the analogous expression valid for a FP equation with only discrete spectrum \cite{risken} to the case in which there also exists a continuous part of the spectrum. 
In order to evaluate the autocorrelation function $R(\tau)= (\langle x(t+\tau)x(t)\rangle-\langle x(t)\rangle^2)/(\langle x^2(t)\rangle-\langle x(t)\rangle^2)$ of the stochastic variable $x(t)$, we make use of the expression 
\begin{equation} 
                \langle x(t+\tau)x(t)\rangle=\sum_{\lambda=\lambda_1}^{\lambda_p}~C^2_{\lambda}e^{-\lambda \tau}+
                                             \int_{\lambda_c}^{+\infty}~C^2_{\lambda}e^{-\lambda \tau} d\lambda, \label{COV}
\end{equation}
where $C_{\lambda} \equiv \int dx~x\,\varphi_{\lambda}(x)$. The analogous of Eq. $(\ref{COV})$ valid for a FP equation with a discrete spectrum is given in Ref. \cite{gardiner}. Eq. $(\ref{COV})$ follows from Eq. $(\ref{W2})$ and from the definition
\begin{eqnarray} 
                \langle x(t+\tau)x(t)\rangle = \iint_{-\infty}^{+\infty} \! dx'\,dx \,
                                               x'\,x\,W_2(x,t;x',t+\tau).  \label{COValt}
\end{eqnarray}
Eq. $(\ref{COV})$ holds true under the assumption that the integrations in $\int dx'\,\int dx$ and $\int d\lambda$ can be interchanged. 

The asymptotic temporal dependence of the autocorrelation
function can have a different behavior depending on the
properties of the eigenvalue spectrum. 
Specifically, we distinguish three different cases:
(a) when the spectrum 
contains the eigenvalue $\lambda_0=0$ and at least one discrete eigenvalue $\lambda_1>0$, 
then $R(\tau)$ decays as $\exp(-\lambda_1 \tau)$.
As an example we mention the Ornstein-Uhlenbeck process 
\cite{ornstein} and the stochastic process associated to the infinite square well quantum potential discussed in
Ref. \cite{risken}; 
(b) when the spectrum contains only one bound state 
corresponding to $\lambda_0 = 0$
and $\lambda_c > 0$ the spectrum has a gap
between the eigenvalue $\lambda_0$  and its 
continuous part. We do not know the general form of the asymptotic behavior of $R(\tau)$ for this class of potentials. 
However, for some specific cases we have found that $R(\tau)$ decays as $e^{-\lambda_c\tau}/\tau^{3/2}$. One example is the V-shaped potential of the FP equation discussed in
Ref. \cite{risken};
(c) when the spectrum contains only one bound state 
corresponding to $\lambda_0 = 0$ and $\lambda_c= 0$ there 
is no gap in the spectrum and a non-exponential asymptotic 
autocorrelation function is in general observed. 
Eq. $(\ref{COV})$ shows that the quantity $\langle x(t+\tau)x(t)\rangle$ is the
weighted sum of an infinite number of exponential  
functions of time, each characterized by a time-scale 
$\lambda^{-1}$. The largest scale of the process is determined by 
the smallest non-vanishing eigenvalue. In case (a) the maximal
time-scale of the process is $\lambda_1^{-1}$. In case (b)
$\lambda_c^{-1}$ is finite and it is the upper bound of the time-scales of the stochastic process. In case (c)
$\lambda_c^{-1}$ diverges and the interval of time-scales of the process is infinite. 

Here we investigate Markovian processes whose maximal time-scale is diverging. This is first done 
by studying the stochastic process associated with a quantum potential $V_S$ given by
\begin{eqnarray}
              V_S=\left \{\begin{array}{ccc}
                                   -V_0~    &~{\rm{if}}~&~|x| \leqslant L, \\
                                   V_1/x^2 ~&~{\rm{if}}~&~|x| >   L,
                   \end{array} \right.  \label{VSchimera}
\end{eqnarray} 
where $L$, $V_0$ and $V_1$ are positive constants. This is an exactly solvable even potential. The parameters $L$, $V_0$ and $V_1$ can be chosen in such a way that the spectrum contains one single discrete eigenvalue $\lambda_0=0$ and a continuous part for $\lambda>0$, as in case (c) discussed above. As a result, the parameters $L$, $V_0$ and $V_1$ are not independent. In fact, the continuity of $\partial_x \psi_0$ in $x=L$ provides a relation between them. The drift coefficient of the corresponding FP equation is
\begin{eqnarray}
              D^{(1)}(x)=\left \{ \begin{array}{cc}
               -2 \sqrt{V_0} \tan (\sqrt{V_0}x) &{\rm{if}}~~|x| \leqslant L ,\\
                 &   \\
                 (1-\sqrt{1+4~V_1})/ x   &{\rm{if}}~~|x| >   L .   
                                  \end{array} \right. \label{D1chimera}
\end{eqnarray}
The associated FP equation describes the dynamics of an overdamped particle moving in a potential that increases logarithmically in $x$.
For $|x| \leqslant L$, the eigenfunction of the ground state is
$\psi_0=B \,\cos(\sqrt{V_0}\,x)$ whereas 
for $|x|>L$ it decays according to $\psi_0=A \,x^{(1-\sqrt{1+4\,V_1})/2}$.
The constants $A$ and $B$ are set by imposing that $\psi_0$ is normalized and continuous in $x=L$. It is worth noting that for $|x|>L$ the stationary pdf $W(x)$ of the
stochastic process is a power-law function decaying as $|x|^{-\alpha}$ 
with $\alpha=\sqrt{1+4\,V_1}-1$.
The normalizability of the eigenfunction of the ground state is ensured if $\alpha>1$.
In the present study we consider stochastic processes with finite variance which implies $\alpha > 3$.
Due to parity arguments, only the odd eigenfunctions $\psi_\lambda^{(odd)}$ of the continuous spectrum give a non-vanishing contribution to $C_\lambda$. For $|x| > L$ the eigenfunction $\psi_\lambda^{(odd)}$ is a linear combination of Bessel functions $\psi_\lambda^{(odd)}=a_\lambda\,\sqrt{x} J_\nu(\sqrt{\lambda}\,x)+b_\lambda\,\sqrt{x}\, Y_\nu(\sqrt{\lambda}\,x)$ where $\nu=(\alpha+1)/2$. For $|x| \leqslant L$ we find $\psi_\lambda^{(odd)}=d_\lambda\,\sin(\sqrt{V_0+\lambda}\,x)$. The coefficients $a_\lambda$, $b_\lambda$ and $d_\lambda$ are fixed by imposing that $\psi_\lambda^{(odd)}$ and its first derivative are continuous in $x=L$ and that $\psi_\lambda^{(odd)}$ are orthonormalized with a $\delta$-function of the energy. By using these eigenfunctions we obtain an exact expression for $C_\lambda$. The further integration required in Eq. $(\ref{COV})$ to obtain $\langle x(t+\tau)\,x(t) \rangle$ cannot be performed analytically. By using Watson's lemma \cite{Olver74} and by considering that the first term of the Taylor expansion of $C_{\lambda}^2$ is proportional to $\lambda^{(\alpha-5)/2}$ for small values of $\lambda$, we find that the asymptotic behavior of $\langle x(t+\tau)\,x(t) \rangle$ valid for large values of $\tau$ is
\begin{equation}
                \langle x(t+\tau)\,x(t) \rangle \sim K\,\tau^{-\beta} ,
\end{equation}
where $K$ is a constant that can be explicitely calculated in terms of the characteristic parameters of the process. The exponent $\beta$ is related to the exponent $\alpha$ through 
\begin{equation}
                \beta=(\alpha-3)/2 .
\end{equation}
To the best of our knowledge this is the first example of a stochastic process which is stationary, Markovian 
and asymptotically power-law autocorrelated.

The above result can also be obtained by considering a different approach based on the 2-point probability density function $W_2$. In this case, the study can be made more general than the previous exact case by considering any quantum potential that asymptotically decays as
\begin{equation}
                V_S(x)\sim \frac{\alpha (\alpha+2)}{4}~
                           \frac{1}{x^{2}} ,
                      \qquad \qquad
                 |x|>\bar x  ,         \label{VSxmenodue}
\end{equation}
and which is characterized by an eigenvalue spectrum as described in case (c). A quantum potential having the above discussed properties has been introduced by two of us in Ref. \cite{lillo}.
The drift coefficient of the FP
equation asymptotically depends on $x$ as $D^{(1)}(x)\sim -\alpha/x$. For $|x| >> \overline{x}$ the eigenfunctions for the class of potentials of Eq. $(\ref{VSxmenodue})$ are the same as those of the potential of Eq. $(\ref{VSchimera})$. 
In general the integration required in Eq. $(\ref{W2})$ cannot be performed analytically. However, the asymptotic behavior of $W_2$ for large values of $\tau$ can be obtained by considering the small energy behavior of the eigenfunctions $\psi_\lambda$. In our calculations we distinguish three regions (i) $|x|<\bar x$, (ii) $\bar x < |x| < x_{\lambda}$ and (iii) $|x| > x_{\lambda}$ where 
$x_{\lambda} \propto \lambda^{-\frac{1}{2}}$.
The asymptotic behavior of $W_2$ for large $\tau$ depends on the values of 
$x$ and $x'$ and the function $\langle x(t+\tau)\,x(t) \rangle$ can be evaluated by splitting the double integrations required in Eq. $(\ref{COValt})$ in eight contributions according to which regions $x$ and $x'$ belong to. 
The eight contributions can be divided into two groups of four. The first group in essentially controlled by the $1/x^2$ region of the quantum potential whereas the second is controlled by the specific form of the potential valid for $|x|<\bar x$.
By using the fact that, for small values of $\lambda$, $\psi_\lambda$ can be approximated as $\psi_\lambda(x) \simeq a_\lambda\,\sqrt{x} J_\nu(\sqrt{\lambda}\,x) \simeq \bigl(2^{(\alpha+2)/2}\,\Gamma((\alpha+3)/2)\bigl)^{-1}\lambda^{(\alpha+1)/4}~x^{1+\alpha/2}$ in region (ii) and $\psi_\lambda(x) \sim \pi^{-1/2}\,\lambda^{-1/4}~\cos(\sqrt{\lambda}\,x + \phi_\lambda)$ in region (iii), with $\phi_\lambda$ a phase factor, we prove that $W_2$ asymptotically behaves as $\tau^{-(\alpha+3)/2}$ and the contributions to $R(\tau)$ associated to the four terms of $W_2$ decay asymptotically as $\tau^{-\beta}$ with $\beta=(\alpha-3)/2$ when both $|x|$ and $|x'|$ are larger than $\bar x$.
We are able to prove this result for $\alpha>4$. This limitation for the $\alpha$ values is an artifact of the approximations used to perform the integrals. In fact, in our calculations we have replaced the cosine terms with unity, thus obtaining an upper bound for the four integrals. The validity of our conclusions for $\alpha>3$ is supported by the exact result obtained for the potential of Eq. $(\ref{VSchimera})$. The contributions to $R(\tau)$ associated to the remaining four terms depend on the specific form of the potential in region (i). For the potential described by Eq. $(\ref{VSchimera})$  these contributions decay to zero faster than $\tau^{-\beta}$. For a generic potential we do not have a general result. However, by considering that the contribution of the remaining four terms is additive, we conclude that $R(\tau)$ decays to zero {\it at least} as $\tau^{-\beta}$.  

We have therefore shown that a class of Markovian stationary processes with a power-law autocorrelation function exists. This is possible because the upper bound for the time-scales
of the process is determined by $\lambda_c^{-1}$ which is infinite for the 
considered processes. However, we wish to stress that the absence of an upper bound for the time-scale is a necessary
but not sufficient condition in order to observe a power-law decaying 
autocorrelation function. 
In fact, let us consider a stochastic process associated with a quantum
potential decaying as $V_S(x)\sim V_1 x^{-\mu}$ and with an eigenvalue spectrum consisting of a single bound state with eigenvalue $\lambda_0=0$ and an attached continuum part, as in case (c). 
The eigenfunction of the ground state is normalizable only for $\mu \leqslant 2$ and when  $0<\mu < 2$ the
asymptotic behavior of $\psi_0$ is proportional to $x^{\mu/4}\exp(-\kappa~x^{1-\mu/2})$, 
where $\kappa=\sqrt{V_1}/(1-\mu/2)$. 
In order to obtain the asymptotic behavior of the autocorrelation function we have to preliminary evaluate the function $C_\lambda$. By using the semi-classical WKB method we estimate the eigenfunctions $\psi_\lambda$. The condition under which this method can be applied is $x^{1-\mu/2} >> \mu/2\,\sqrt{V_1}$, and is fulfilled for any $0<\mu<2$. The WKB method gives two approximations of the eigenfunctions that hold in the regions $x<x_\lambda$ and $x>x_\lambda$, where $x_\lambda$ is here the turning point defined by $V_1 x_\lambda^{-\mu}=\lambda$. These approximations can be used to obtain the asymptotic behavior of $C_{\lambda}$ valid for small values of $\lambda$. Specifically, we obtain
\begin{eqnarray}
           &&  \hspace{-0.35 in} 
               C_{\lambda} \simeq 2\,(2+\mu)^{2 \over{2+\mu}}\,
                                   \Gamma \left({{4+\mu}\over{2+\mu}}\right)\,
                                   {{V_1^{1 \over{2+\mu}}} 
                                       \over
                                    {\lambda^{{{4+\mu}\over{2+\mu}}}}}\,
                                    \exp \left(
                                               {-{\eta \over{\lambda^{{1\over \mu} - {1 \over 2}}}}} 
                                         \right) ,  \\
          && \hspace{-0.35 in} 
               \eta = \sqrt{\pi}~
                   {{2 \mu}\over{4-\mu^2}}~
                   V_1^{1/\mu}~
                   \Gamma \left({{3}\over{2}}+{{1}\over{\mu}}\right)/
                   \Gamma \left(1+{{1}\over{\mu}}\right) .  \nonumber                            
\end{eqnarray}  
By applying the saddle point method to the integration of Eq. $(\ref{COV})$, we find for the asymptotic behavior of the autocorrelation function the expression
\begin{eqnarray}
           &&   R(\tau)\sim R_\infty\,\tau^{\frac{5 \mu^2 + 24 \mu -4}{2(\mu+2)^2}}~
                            \exp \left({-\xi~\tau^{\frac{2-\mu}{2+\mu}}} \right) , \\
           &&   \xi = (2\,\eta)^{{2 \mu}\over{2 + \mu}}~
                      \left(
                            \Bigl( {{2-\mu}\over{2 \mu}} \Bigl)^{{2 \mu}\over{2+\mu}}~+~
                            \Bigl( {{2 \mu}\over{2-\mu}} \Bigl)^{{2-\mu}\over{2+\mu}}  
                      \right) ,  \nonumber                  
\end{eqnarray}
where $R_\infty$ is a constant. The asymptotic behavior of the autocorrelation function is therefore dominated by a stretched exponential function. This is observed in spite of the fact that the upper bound of time-scales is infinite as for the processes with $\mu=2$. This is due to the fact that, despite the presence of an infinite range of time-scales, the weight of the longest time-scales (given by $C^2_{\lambda}$ for $\lambda \to 0$) is too small to give a power-law decay of the autocorrelation function. This is seen by considering that we can write the autocorrelation function
for the considered processes as
\begin{equation}
                R(\tau)=\int_0^{+\infty} \pi(\tau_s) e^{-\tau/\tau_s} ~ d\tau_s ,
\end{equation}
i.e. as the weighted sum of exponentially decaying functions with
characteristic time-scale $\tau_s =\lambda ^{-1}$ and weights
$\pi(\tau_s)=\tau_s^{-2} C^2_{\tau_s}/\sigma^2$, where $\sigma$ is
the standard deviation of the stationary pdf.
For power-law correlated processes ($\mu=2$) the weights decay as a
power-law proportional to $1/\tau_s^{\beta+1}$, whereas for
$0<\mu<2$ one has a decay proportional to $\tau_s^{\frac{4}{2+\mu}} 
\exp(-2\eta \tau_s^{\frac{2-\mu}{2\mu}})$. In the last case the weight
of the largest time-scales decay too fast to zero to ensure a power-law autocorrelation
function. We also observe that
\begin{equation}
                  \langle \tau_s\rangle=
                  \int_0^{+\infty} \pi(\tau_s) \tau_s ~ d\tau_s=
                  \int_0^{+\infty} R(\tau)~ d\tau .
\end{equation}
Hence the integral of the autocorrelation function is a measure
of the mean characteristic time of the process. This mean characteristic time $\langle \tau_s\rangle$
diverges only for the class of stochastic processes associated with a quantum
potential decaying as $x^{-2}$ for $|x| \to \infty$ and with an eigenvalue spectrum as in case (c) when $0<\beta\leqslant 1$. 

Let us finally consider the stochastic process $s(t)=\int_0^t dt' x(t')$. It can be considered as the displacement process associated to the velocity process $x(t)$. The mean squared displacement $\langle \Delta s^2(t) \rangle$ is given by
\begin{eqnarray}
                \langle \Delta s^2(t) \rangle = 2\int_0^t    dt' \,
                                         \int_0^{t'} d\tau \,
                                         \langle x(t+\tau)\,x(t) \rangle .
\end{eqnarray} 
For any stochastic process $x(t)$ with autocorrelation function that asymptotically decays proportionally to $\tau^{-\beta}$ the temporal dependence of the mean squared displacement 
$\langle \Delta s^2(t)\rangle$ for large values of $t$
strongly depends on $\beta$. Specifically: 
(i) when $\beta>1$, then $\langle \Delta s^2(t) \rangle \sim D_\gamma\,t$, i.e. $s(t)$ is a diffusive stochastic process;
(ii) when $\beta=1$ one has $\langle \Delta s^2(t) \rangle \sim D_\gamma\,t\,\ln(t)$;
(iii) when $0<\beta<1$ the process becomes superdiffusive, i.e. $\langle \Delta s^2(t)\rangle \sim D_\gamma\,t^{2-\beta}$. The case when the displacement process $s(t)$ is diffusive corresponds to an integrable autocorrelation function of $x(t)$. Conversely, when $R(\tau)$ is not integrable the process $s(t)$ is not diffusive. 
The spectral density $S(f)$ of the process $x(t)$ is $S(f) \propto 1/f^{1-\beta}$ 
for low frequency values. 
Hence, for values of $\beta$ slightly larger than zero the 
process $x(t)$ is characterized by a $1/f$ spectral density at low frequency. 
By introducing a long-range correlated stationary Markovian process we have shown that an $1/f$-like stochastic 
process is described by a FP equation with a constant diffusion coefficient and with a
drift coefficient which depends asymptotically on $x$ as $-\alpha/x$
when $\alpha$ is slightly larger than 3. One 
example is the drift $D^{(1)}(x)$ of Eq. $(\ref{D1chimera})$, for 
which the nonlinear Langevin equation is
$\dot x(t)=D^{(1)}(x) + ~\Gamma (t)$,
where $\Gamma (t)$ is a Gaussian white noise of zero mean 
and unit variance. This is an example of a nonlinear Langevin equation describing a $1/f$ noise which is a stationary, Markovian and non-Gaussian process with finite variance.

In summary, we show that there exist stationary Markovian 
processes which are characterized by an interval of time-scales ranging from a finite value to infinite.  
We present analytical evidence that when (i) the range of time-scale is
infinite and (ii) the mean time-scale $\langle \tau_s\rangle$
of the process diverges the Markovian process is 
long-range correlated and has associated a low-frequency
region of the spectral density which may be $1/f$-like. 
We show analytically that both requirements occur for processes
described by a FP equation with constant
diffusion coefficient which has associated
a quantum potential $V_S(x)$ asymptotically proportional to $x^{-2}$ and with an eigenvalue spectrum consisting of a single discrete eigenvalue $\lambda_0=0$ and an attached continuum part. 
We also show that the divergence of the maximal time-scale 
of the process is necessary but not sufficient to imply the
presence of long-range correlation. What is crucial to 
observe long-range correlation
is that the mean time-scale $\langle \tau_s\rangle$
diverges. 

We thank INFM (FRA and PAIS projects), ASI and MIUR for financial support.

\end{document}